\documentclass[prd,aps,twocolumn,showkeys,nofootinbib]{revtex4-2}

\usepackage{amsmath}
\usepackage{amsfonts}
\usepackage{amssymb}	
\usepackage{graphicx}
\usepackage{bm}
\usepackage{color}
\usepackage{ulem}
\usepackage{commath}
\usepackage[T1]{fontenc}
\usepackage{lmodern}
\usepackage{xcolor}
\usepackage{hyperref}
\allowdisplaybreaks

\def\Msun{M_{\odot}}
\def\GMc2{G M_{\odot} c^{-2}}

\def\lm{{\ell m}}
\def\EOB{{\rm EOB}}
\def\GSF{{\rm GSF}}

\def\lm{{\ell m}}

\def\lm{{\ell m}}

\newcommand\be{\begin{equation}}
\newcommand\ee{\end{equation}}

\def\Msun{M_\odot}
\def\Qo{Q_{\omega}}

\newcommand{\TEOBResumS}[1]{\texttt{TEOBResumS{#1}}}

\def\Dali{\TEOBResumS{-Dalí}}
\def\Giotto{\TEOBResumS{-GIOTTO}}

\begin{document}

\title{
  Effective-one-body waveforms for extreme-mass-ratio binaries: \\
  Consistency with second-order gravitational self-force quasicircular results \\
  and extension to nonprecessing spins and eccentricity}

\author{Angelica \surname{Albertini}${}^{1,2}$}
\author{Rossella \surname{Gamba}${}^{3}$}
\author{Alessandro \surname{Nagar}${}^{4,5}$}
\author{Sebastiano \surname{Bernuzzi}${}^{3}$}

\affiliation{${}^1$Astronomical Institute of the Czech Academy of Sciences,
Bo\v{c}n\'{i} II 1401/1a, CZ-141 00 Prague, Czech Republic}
\affiliation{${}^2$Faculty of Mathematics and Physics, Charles University in Prague, 18000 Prague, Czech Republic}
\affiliation{${}^3$Theoretisch-Physikalisches Institut, Friedrich-Schiller-Universit{\"a}t Jena, 07743, Jena, Germany}
\affiliation{${}^4$INFN Sezione di Torino, Via P. Giuria 1, 10125 Torino, Italy} 
\affiliation{${}^5$Institut des Hautes Etudes Scientifiques, 91440 Bures-sur-Yvette, France}

\begin{abstract}
We present a first complete implementation of an effective-one-body
(EOB) model for extreme-mass-ratio inspirals (EMRIs) that incorporates 
aligned spins (on both the primary and the secondary)
as well as orbital eccentricity.
The model extends \Dali{} for these binaries  by
(i) recasting conservative first-order gravitational self-force (1GSF) information
in the resummed EOB potentials;
(ii) employing a post-Newtonian (PN) $3^{+19}$PN-accurate (3PN comparable-mass terms
hybridized with test-particle terms up to 22PN relative order)
expression for the gravitational-wave flux at infinity;
(iii) using an improved implementation of the horizon flux that better approximates its test-mass representation.
With respect to our previous work [Phys.~Rev.~D 106 (2022) 8, 084062],
we demonstrate that the inclusion of the $3^{+19}$PN-accurate $\ell=9$
and $\ell=10$ modes in the flux at infinity significantly improves the
model's agreement with second-order accurate GSF (2GSF)
circular waveforms. 
For a standard EMRI with mass ratio $q \equiv m_1/m_2 = 5 \times 10^4$
and $m_2 = 10 M_\odot$, the accumulated EOB/2GSF dephasing is
$\lesssim $~rad for $\sim 1$~yr of evolution, which is consistent with
the standard accuracy requirements for EMRIs.  
We also showcase the generation of eccentric and spinning waveforms
and discuss future extensions of our EOB towards a physically
complete model for EMRIs.
\end{abstract}

\date{\today}

\maketitle

\section{Introduction}

After the first detection of gravitational waves, the LIGO-Virgo-KAGRA Collaboration 
witnessed many more black hole binary coalescences, mostly related to stellar-mass black hole binaries.
The forthcoming generation of gravitational-wave detectors will instead provide us with signals from
a more diverse collection of sources, among which are extreme-mass-ratio inspirals (EMRIs).
These are systems made of a stellar-mass compact object slowly inspiralling into a black hole with
mass $M \sim 10^4 - 10^7 M_\odot$, thus emitting a waveform that encodes the features of the background
space-time and allows to precisely infer astrophysical properties of the source and to test general relativity (GR).
However, to actually be able to detect and analyze this kind of waveforms, we need waveform templates that
are physically complete, accurate and fast to evaluate.

Given its flexibility, the effective-one-body (EOB) approach~\cite{Buonanno:1998gg,Buonanno:2000ef,Damour:2000we,Damour:2001tu,Damour:2015isa} 
is the natural framework to provide solutions to this problem that are {\it complementary to} and partly informed 
by~\cite{Damour:2009sm} gravitational-self-force results~\cite{Barack:2018yvs}. 
A recent work~\cite{Nagar:2022fep} precisely showed that it is natural, within 
the EOB framework, to describe all physical effects that are expected to shape the waveform of a radiation-reaction 
driven EMRI by considering together spins and eccentricity. The model of Ref.~\cite{Nagar:2022fep} is essentially a
modification of the \Dali{} model that includes the complete first-order self-force (1SF) information in the 
Hamiltonian (i.e. the conservative sector), but keeps the other elements the same as in the implementation that is valid for comparable-mass
binaries~\cite{Nagar:2021gss,Nagar:2021xnh,Bonino:2022hkj}. This in particular concerns the amount of 
analytical information incorporated in the azimuthal radiation reaction force that drives the inspiral. 
Its state of the art is described in Ref.~\cite{Nagar:2020pcj} and is such that, when restricted to the limit of a test-particle
orbiting a Schwarzschild black hole on circular orbits (i.e. at zeroth post-adiabatic order, or 0PA), 
it gives an approximation to the corresponding exact fluxes~\cite{Cutler:1993vq} that might be insufficient 
to accurately drive an EMRI inspiral. The recent development of the first waveform model with second-order
gravitational-self-force information (2GSF)~\cite{Wardell:2021fyy} allowed for the possibility to benchmark the model for 
quasi-circular nonspinning binaries in a regime of mass ratios that at the moment cannot be efficiently 
and systematically covered by numerical relativity (NR) simulations 
(see however Refs.~\cite{Lousto:2022hoq,Nagar:2022icd,Rosato:2021jsq,Lousto:2020tnb} for 
ground-breaking studies that push NR simulations up to mass ratios 1000:1). 
In particular, Refs.~\cite{Albertini:2022rfe,Albertini:2022dmc} clearly showed that the EOB fluxes
of Ref.~\cite{Nagar:2020pcj} had to be improved to yield a closer approximation to the 0PA fluxes 
driving the 2GSF evolution of Ref.~\cite{Wardell:2021fyy}. Ref.~\cite{Albertini:2022dmc} 
obtained a good mutual consistency between 2GSF and EOB evolutions for quasi-circular, nonspinning, 
large-mass-ratio binaries thanks to: 
(i) the GSF-tuned potentials presented in Ref.~\cite{Nagar:2022fep} (see also Ref.~\cite{Antonelli:2019fmq});
(ii) the use of hybrid 3PN comparable-mass terms and 22PN test mass terms in the (factorized and resummed~\cite{Damour:2008gu,Nagar:2016ayt,Messina:2018ghh}) multipoles 
in the infinity flux (up to $\ell = 8$)~\cite{Fujita:2012cm} -- henceforth denoted as $3^{+19}$PN-accurate -- instead of the (mostly) $3^{+3}$PN-accurate ones~\cite{Nagar:2020pcj}; 
(iii) an improved implementation of the horizon flux that better approximates its test-mass exact representation.
Here we show that the inclusion of the $\ell=9$ and $\ell=10$ multipolar contributions at $3^{+19}$PN accuracy 
to the EOB infinity flux is sufficient to remove the remaining few radians of EOB/GSF phase differences for a standard EMRI
pointed out in Fig.~7 of Ref.~\cite{Albertini:2022dmc}.
We finally give some details about the performance of the public \TEOBResumS{} infrastructure.

This paper is organized as follows: in Sec.~\ref{sec:eob} we collect the elements of \TEOBResumS{} 
that are modified and evaluate their impact on the phasing. In Sec.~\ref{sec:LISA} we recast the results for a typical EMRI and discuss the 
performance of the public implementation. Our final remarks are gathered in Sec.~\ref{sec:end}.
We define the mass ratio $q \equiv m_1/m_2 \ge 1$,  the symmetric 
mass ratio $\nu\equiv m_1 m_2/M^2$, where $m_{1,2}$ are the masses of the two 
bodies, $M\equiv m_1+m_2$ and we use the convention $m_1\geq m_2$. 
The dimensionless spin variables are denoted as $\chi_{1,2}\equiv S_{1,2}/(m_{1,2})^2$, where $(S_1,S_2)$ are the individual, dimensionful, spin components along the
direction of the orbital angular momentum. We also use the effective spin $\chi_{\rm eff}= \chi_1 m_1 / M + \chi_2 m_2 / M$.
We use units with $G=c=1$ and typically normalize quantities by the total mass $M$, e.g. time is $t\equiv T/M$, radial separation $r\equiv R/M$ etc.
The content of this paper strongly builds upon Refs.~\cite{Albertini:2022rfe,Albertini:2022dmc}, and we assume the reader to be familiar with them.

\section{Tailoring the EOB model to EMRIs}
\label{sec:eob}

To orient the reader, we quickly recall the basic elements of the EOB approach~\cite{Buonanno:1998gg,Buonanno:2000ef,Damour:2000we,Damour:2001tu,Damour:2015isa},
while we address to Ref.~\cite{Albertini:2022dmc} and references therein for additional details.  
The EOB Hamiltonian is found by mapping the two-body problem in GR into the motion of a single body with the reduced mass of the system moving in an effective metric, 
it is written as a function of the EOB potentials and of the dynamical variables and encodes the conservative dynamics. The latter yields equations of motion that are complemented by the radiation reaction, accounting for the amount of energy and angular momentum carried away
by gravitational waves, both towards infinity and into the black hole horizons.
The EOB model for EMRIs currently relies on (i) the EOB potentials of Ref.~\cite{Nagar:2022fep}, which are at linear order in $\nu$ and
their 1SF term includes a fit to GSF data\footnote{We note that due to the gauge choice, these potentials are singular at the light-ring, hence limiting this
version of the model to inspiral-plunge templates.}; (ii) the $3^{+19}$PN resummed infinity flux as discussed in Sec.~IIB of Ref.~\cite{Albertini:2022dmc}; (iii) an improved implementation of the horizon flux. 

The gravitational strain waveform is written
as a multipole expansion in $s=-2$ spin-weighted spherical harmonics
\be
h_+ - i h_\times = \dfrac{1}{D_L}\sum_{\ell=2}^{\ell_{\rm max}} \sum_{m=-\ell}^{\ell}h_\lm\,{}_{-2}Y_\lm(\iota,\Phi) \ ,
\ee
where $D_L$ indicates the luminosity distance and $\iota$ ($\Phi$) is the polar (azimuthal) angle of the line of sight
with respect to the orbital plane. We will only present results involving the
dominant quadrupole multipole, that is decomposed in phase and amplitude as
\be
\label{eq:RWZnorm}
h_{22} (t)= A(t) e^{-i \phi(t)}.
\ee
The instantaneous
gravitational wave frequency (in units of $M$) is defined as $\omega\equiv \dot{\phi}$.
Within the usual EOB scheme~\cite{Damour:2008gu} each multipole is factorized as 
$h_\lm = h^{\rm Newt}_\lm \hat{h}_\lm$, where the first contribution is the leading, 
Newtonian one, and the PN corrections is similarly written in factorized form as
\be
\label{eq:hlm}
\hat{h}_\lm = \hat{S}^{(\epsilon)}_{\rm eff} \, T_\lm e^{i \delta_\lm}  \left( \rho_\lm \right)^{\ell} ,
\ee
where $\epsilon=(0,1)$ is the parity of $\ell + m$, $\hat{S}^{(\epsilon)}_{\rm eff}$ is the effective 
source of the field (effective energy or Newton-normalized angular momentum depending on the parity
of the mode~\cite{Damour:2008gu}), $T_\lm$ is the tail factor, which resums an infinite number of
leading-order logarithms, while $\rho_\lm$ and $\delta_\lm$ are the residual amplitude and phase
corrections, respectively. The $m = 0$ mode is absent, as its effect is considered negligible for 
a quasi-circular inspiral. From the waveform strain modes one computes the flux modes entering
the azimuthal radiation force that drives the quasi-circular inspiral yielding the loss of mechanical angular
momentum. Following for example Ref.~\cite{Damour:2012ky}, Hamilton's equation for the 
angular momentum reads
\be
\dot{p}_\varphi = \hat{\cal F}_{\varphi} \ ,
\ee
where the radiation reaction force  $ \hat{\cal F}_{\varphi}$ is given by
\be 
\hat{\cal F}_{\varphi}=-\dfrac{32}{5} r_\omega^4 \Omega^5 \hat{f}(v_\varphi^2;\nu) \ ,
\ee
where $\Omega$ is the orbital frequency, $r_\omega$ is a radial separation defined in such a way that it still fulfils Kepler’s law
 during the late-inspiral and plunge~\cite{Damour:2006tr,Damour:2007xr},
$v_\varphi = r_\omega \Omega$ and $\hat{f}(v_\varphi^2;\nu)$ is the reduced flux function. 
This is given by the total energy flux, summed over a certain number of multipoles, both for
the flux at infinity and at the horizon~\cite{Damour:2012ky}. In particular, for the quasi-circular 
flux at infinity, we have that $\hat{f}^\infty \equiv F^{\infty}/F_{22}^{\rm Newt}$, where
\be
F^{\infty}\equiv \sum_{\ell=2}^{\ell_{\rm max}}\sum_{m=-\ell}^{\ell} F_\lm^{\rm Newt} |\hat{h}_\lm |^2 .
\ee
In Ref.~\cite{Albertini:2022dmc} we used $\ell_{\rm max}=8$ and the $\rho_{\lm}$ functions of 
Eq.~\eqref{eq:hlm} at $3^{+19}$PN order. Here we use the {\it same} global PN accuracy, so that we
have 22PN-accurate $\rho_\lm$'s  for $\nu=0$ (see Ref.~\cite{Fujita:2012cm}), but we 
fix $\ell_{\rm max}=10$.  By contrast, the description of the horizon flux is precisely the one 
presented in Ref.~\cite{Albertini:2022dmc}.

%==========
% Qomg 50000
%==========
\begin{figure}[t]
\center
\includegraphics[width=0.4\textwidth]{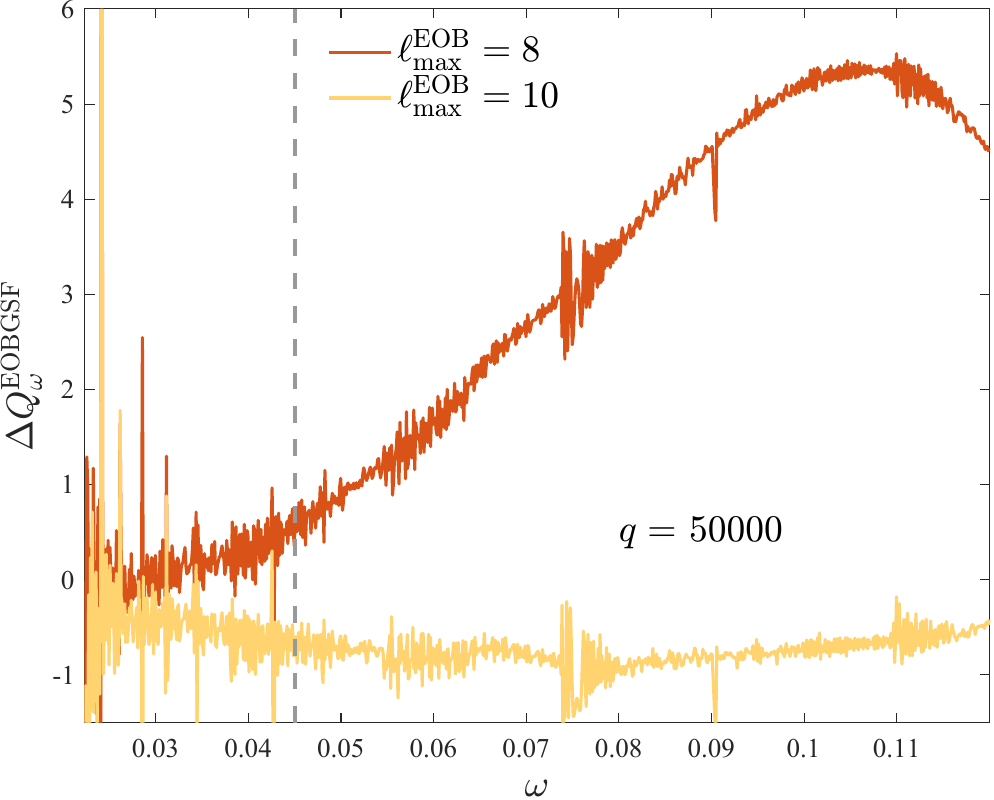} 
\caption{\label{fig:DQo1}Probing how the EOB/GSF difference in $\Qo$ for $q=50000$ changes when
adding more multipoles to the EOB flux. The red line is related to the same model used in Ref.~\cite{Albertini:2022dmc}
(see the lower panel in Fig.~7 therein), where the EOB infinity flux included modes up to $\ell =8$,
and corresponds to an integrated phase difference $\Delta\phi_{\ell_{\rm max} = 8}^{\rm EOBGSF} = 3.1$ rad.
The yellow line corresponds to the addition of the $\ell = 9,10$ modes into the infinity flux, and lowers the integrated
phase difference to $\Delta\phi_{\ell_{\rm max} = 10}^{\rm EOBGSF} = -1.1$ rad. When considering the frequency 
interval $\omega = [0.045, 0.12]$, i.e. from the dashed grey line until the end, the integrated phase differences are respectively
$\Delta\phi_{\ell_{\rm max} = 8}^{\rm EOBGSF} = 3.0$ and $\Delta\phi_{\ell_{\rm max} = 10}^{\rm EOBGSF} = -0.7$.
This second interval corresponds to more physically meaningful EMRI parameters (see the text).}
\end{figure}

\subsection{Effects on phasing}

In Refs.~\cite{Albertini:2022rfe,Albertini:2022dmc} we compared the EOB waveforms to the 2GSF ones from Ref.~\cite{Wardell:2021fyy} and 
assessed the phasing accuracy in terms of the gauge-invariant adiabaticity parameter\footnote{For technical reasons we evaluate this function with 
waveforms computed with the private {\tt MATLAB} implementation of the code, that allows for less noise in the phase acceleration $\dot{\omega} = \ddot{\phi}$.
We have however checked the phase consistency between the {\tt MATLAB} and {\tt C} evolutions.}
\be
\Qo \equiv \frac{\omega^2}{\dot{\omega}},
\ee
that is computed for both the EOB and GSF waveforms. This quantity can also be expanded in powers 
of $\nu$ and the coefficients can be evaluated at fixed values of $\omega$ (see below).
The accumulated phase difference between EOB and GSF  in the frequency interval $(\omega_1,\omega_2)$ 
is then given by the integral
\begin{align}
\Delta\phi^{\rm EOBGSF} &\equiv \phi^{\EOB} - \phi^{\GSF} \\ 
									     &= \int_{\omega_1}^{\omega_2} \Delta \Qo^{\rm EOBGSF} d\log\omega \ ,
\end{align} where $\Delta \Qo^{\rm EOBGSF}\equiv \Qo^{\rm EOB}-\Qo^{\rm GSF}$.
We consider an EMRI with $q=50000$ in the frequency range $(\omega_1,\omega_2)=(0.0224, 0.12)$, i.e. the same
binary studied in Ref.~\cite{Albertini:2022dmc}.
In Fig.~\ref{fig:DQo1} we compare  the difference $\Delta\phi^{\rm EOBGSF}$  obtained with $\ell_{\rm max}=8$ 
with the new one obtained with $\ell_{\rm max}=10$. One sees the visual improvement in this second case
with $\Delta \Qo^{\rm EOBGSF}$ that does not grow secularly as before. Quantitatively, the phase difference 
$\Delta\phi_{\ell_{\rm max} = 8}^{\rm EOBGSF} = 3.1$ drops to $\Delta\phi_{\ell_{\rm max} = 10}^{\rm EOBGSF} = -1.1$.
In Ref.~\cite{Albertini:2022dmc} we proved how the use of the $3^{+19}$PN flux at infinity and of the
improved implementation of the horizon flux were crucial in increasing the agreement with GSF, especially for $q = 50000$.
While being a minor upgrade compared to these more fundamental modifications, the addition of the $\ell = 9,10$ modes is effective
in flattening the $\Qo$ difference, thus proving that for EMRIs it is sufficient to include modes up to $\ell = 10$ in the radiation reaction of our model.
%we view the addition of  the 
%$\ell = 9,10$ multipoles to the infinity flux as the icing on the cake that finally flattens the phase difference.

%===========
% Qomg 1,2,3
%===========
\begin{figure}[t]
\center
\includegraphics[width=0.4\textwidth]{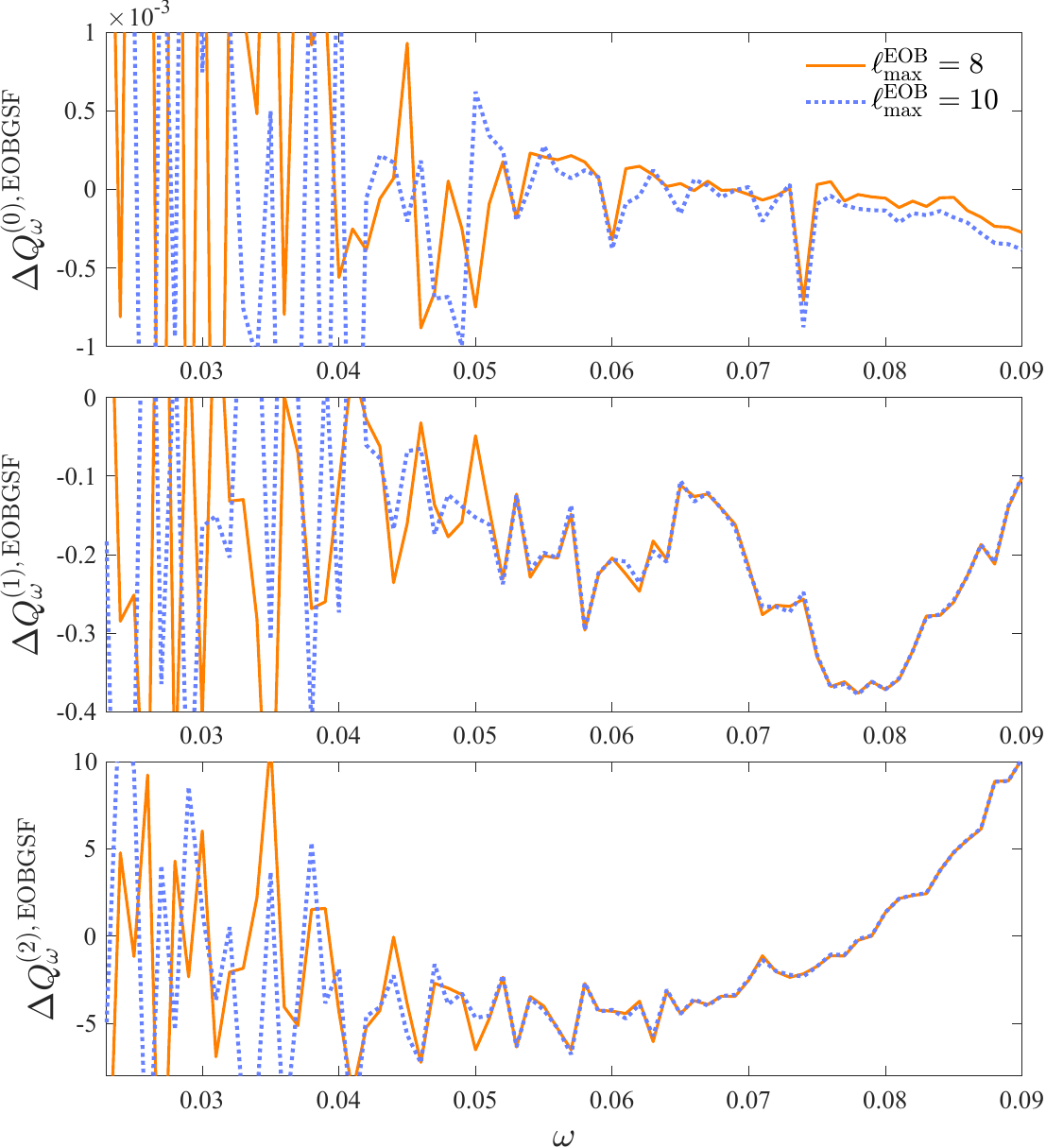} 
\caption{\label{fig:diffQ012} EOB/GSF difference in $\Qo^{(0)},\Qo^{(1)},\Qo^{(2)}$ both with the EOB flux using up to $\ell = 8$ and up to $\ell = 10$.
The lowered difference in $\Qo^{(0)}$ further explains why the $\ell_{\rm max} = 10$ model better agrees with GSF results.}
\end{figure}

To gain further insight on the importance of the various contributions, we follow Ref.~\cite{Albertini:2022dmc}
and write $\Qo$ as an expansion in $\nu$ as
\be
\label{eq:Qomg}
\Qo = \frac{\Qo ^{(0)}}{\nu} + \Qo ^{(1)} +  \nu \, \Qo ^{(2)} ,
\ee
where the superscripts indicate the PA order. As for GSF, an analytical evaluation of 
the $\Qo^{(i)}$ coefficients can be found in Appendix B of Ref.~\cite{Albertini:2022rfe}, 
while for EOB it was performed for circular orbits in Sec.~IV of Ref.~\cite{Albertini:2022dmc},
see Eqs.~(23)-(25) therein. Following these two expansions, for circular orbits the 0PA term depends only on the
dissipative piece of the first-order self force. The 1PA term also depends (i) on the conservative piece of the
fist-order self force, which in EOB language translates to the linear-in-$\nu$ contribution to the main 
potential $A$, (ii) on the dissipative piece of the second-order self force, which in EOB language translates 
to the linear-in-$\nu$ contributions to the $\rho_\lm$'s as well as other effects coming from the effective
source $\hat{S}^{(\epsilon)}_{\rm eff}$ and the tail terms $T_\lm$ (see also Ref.~\cite{vandeMeent:2023ols}).
The 2PA term further depends on the conservative second-order self force ($\nu^2$ term into $A$) and on the 
dissipative third-order self force ($\nu^2$ terms in the $\rho_\lm$'s). However, as for EOB there are more
contributions that come into play when one considers all the non-circular information contained in the model. Specifically,
(i) other two EOB potentials, $D$ and $Q$, enter the Hamiltonian when the radial momentum $p_{r_*}$ is nonzero; 
(ii) the $\rho_\lm$'s are evaluated as functions of $x = v_{\varphi}^2  = (r_\omega \Omega)^2$,
where $r_\omega$ accounts for non-circularity during the late inspiral and plunge, as mentioned before.

We follow Ref.~\cite{Albertini:2022dmc} and determine the $\Qo^{(i)}$'s
via a suitable fitting procedure. In Ref.~\cite{Albertini:2022dmc} we used
mass ratios $q=\{26,32,36,50,64,128,500\}$. Here, for both models, we include instead
$q = \{ 26, 32, 64, 128, 500, 5000, 50000 \}$, so that the larger mass ratios are also
contributing to the fit. 
 The EOB-GSF differences in the $\Qo^{(i)}$ coefficients is plotted in Fig.~\ref{fig:diffQ012}, where
\be
\Delta \Qo^{(i), \rm EOBGSF} = \Qo ^{(i), \rm EOB} -  \Qo ^{(i), \rm GSF}, \;\; i = 0,1,2 \, .
\ee
As a consistency check, we see that the differences in $\Qo ^{(1)}$ and $\Qo ^{(2)}$ are not affected 
by the addition of the $\ell = 9,10$ multipoles, while the difference in $\Qo ^{(0)}$ decreases. 
Given the analytical form of the $\Qo$ expansion, the 0PA coefficient
has a larger impact on the dephasing as the mass ratio $q$ increases ($\nu$ decreases), so that the lowered difference 
in $\Qo ^{(0)}$ further explains the improved agreement we reach with this simple modification of the EOB flux.

\section{Perspectives for LISA: parameter space, accuracy and performance}
\label{sec:LISA}

%=============
% Figure benchmark
%=============
\begin{figure}[t]
	\center
	\includegraphics[width=0.5\textwidth]{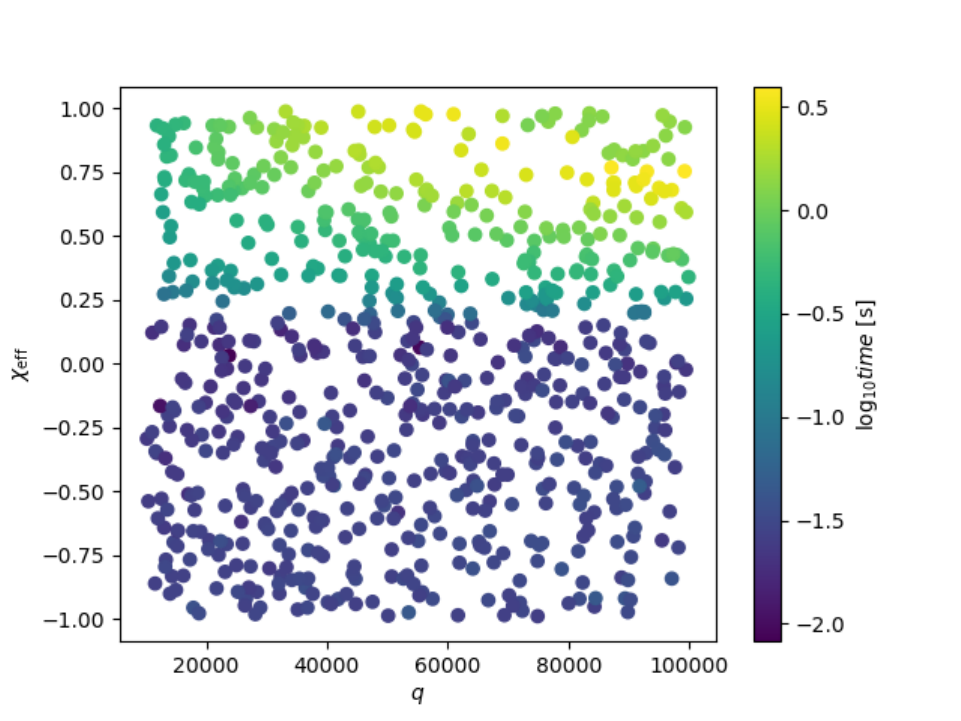} 
	\includegraphics[width=0.5\textwidth]{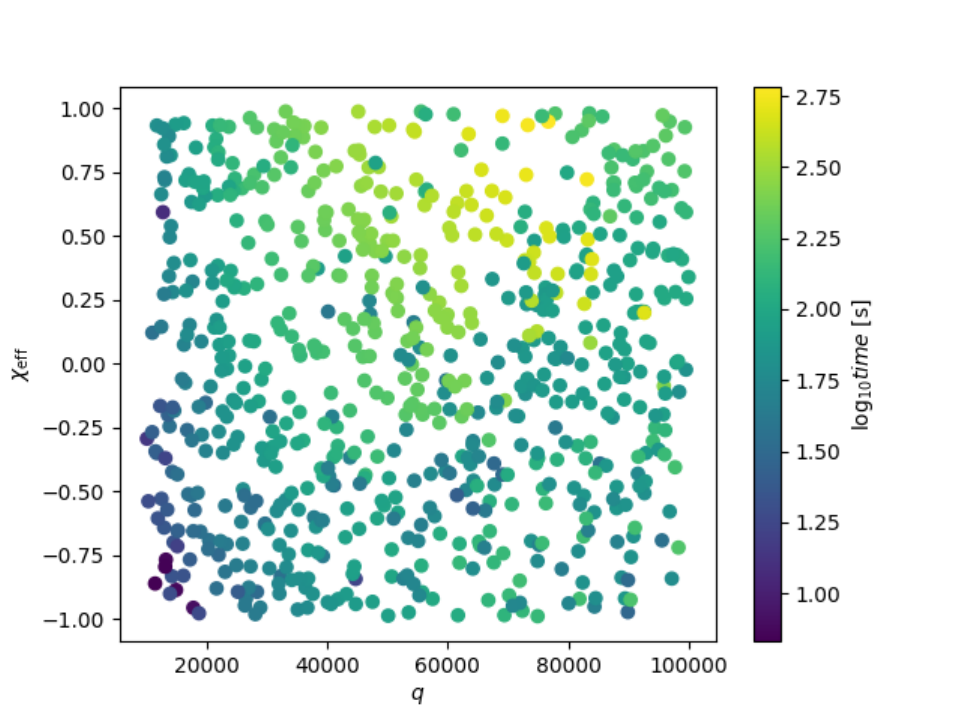} 
	\caption{\label{fig:benchmark}Benchmark of the quasi-circular model for systems with total mass $M=10^7\Msun$ 
	from initial frequency $10^{-4}$~Hz. We display the time necessary to generate a waveform up to $r/M=5$ for 
	$\sim 700$ binary configurations with varying mass ratios $q$ and effective spin $\chi_{\rm eff}= \chi_1 m_1 / M + \chi_2 m_2 / M$, 
	with or without the iterated post-adiabatic approximation~\cite{Nagar:2018gnk} (top and bottom panels, respectively).}
\end{figure} 

%=============
% Figure r PA
%=============
\begin{figure}[t]
	\center
	\includegraphics[width=0.5\textwidth]{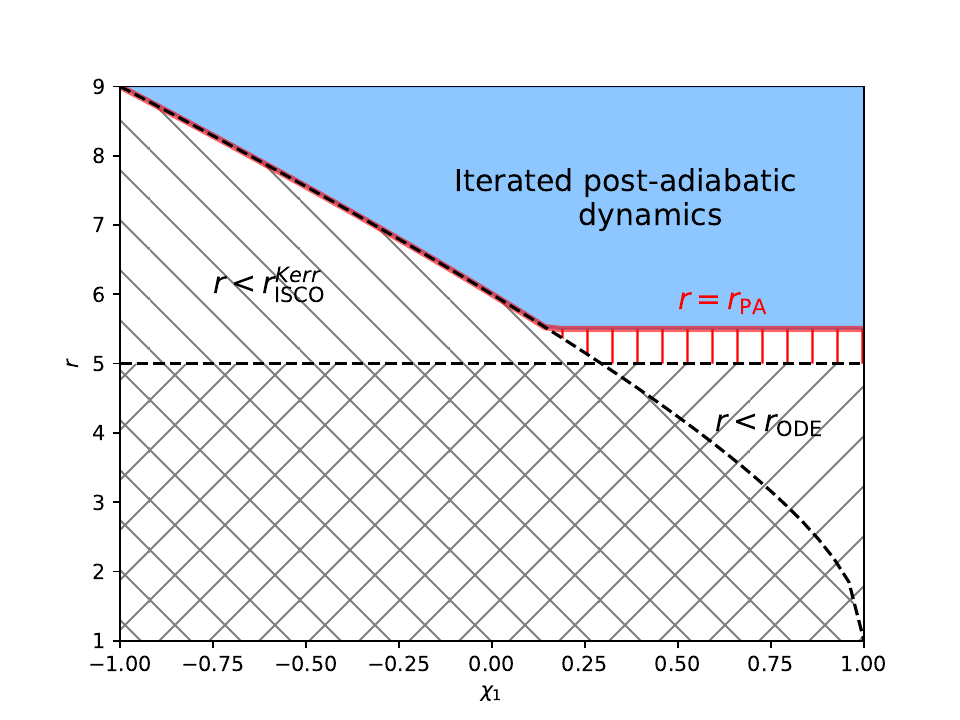} 
	\caption{\label{fig:rbounds}Illustrative depiction of the boundaries we impose to determine the value of $r_{\rm PA}$ 
	at which we terminate the iterative post-adiabatic evolution. Drawing inspiration from the dynamics of a test
	particle on a Kerr black hole, we evolve the system iteratively until either we reach the innermost circular orbit (ISCO) 
	$r = r_{\rm ISCO}^{\rm Kerr}$ or when we hit the fixed threshold $r = 5.5$ (red line in plot). At that point, the evolution is
	continued numerically up to $r =  r_{\rm final} = 5$.}
\end{figure}

%=============================
% Figure eccentric spinning, multipoles
%=============================
\begin{figure}[t]
	\includegraphics[width=0.48\textwidth]{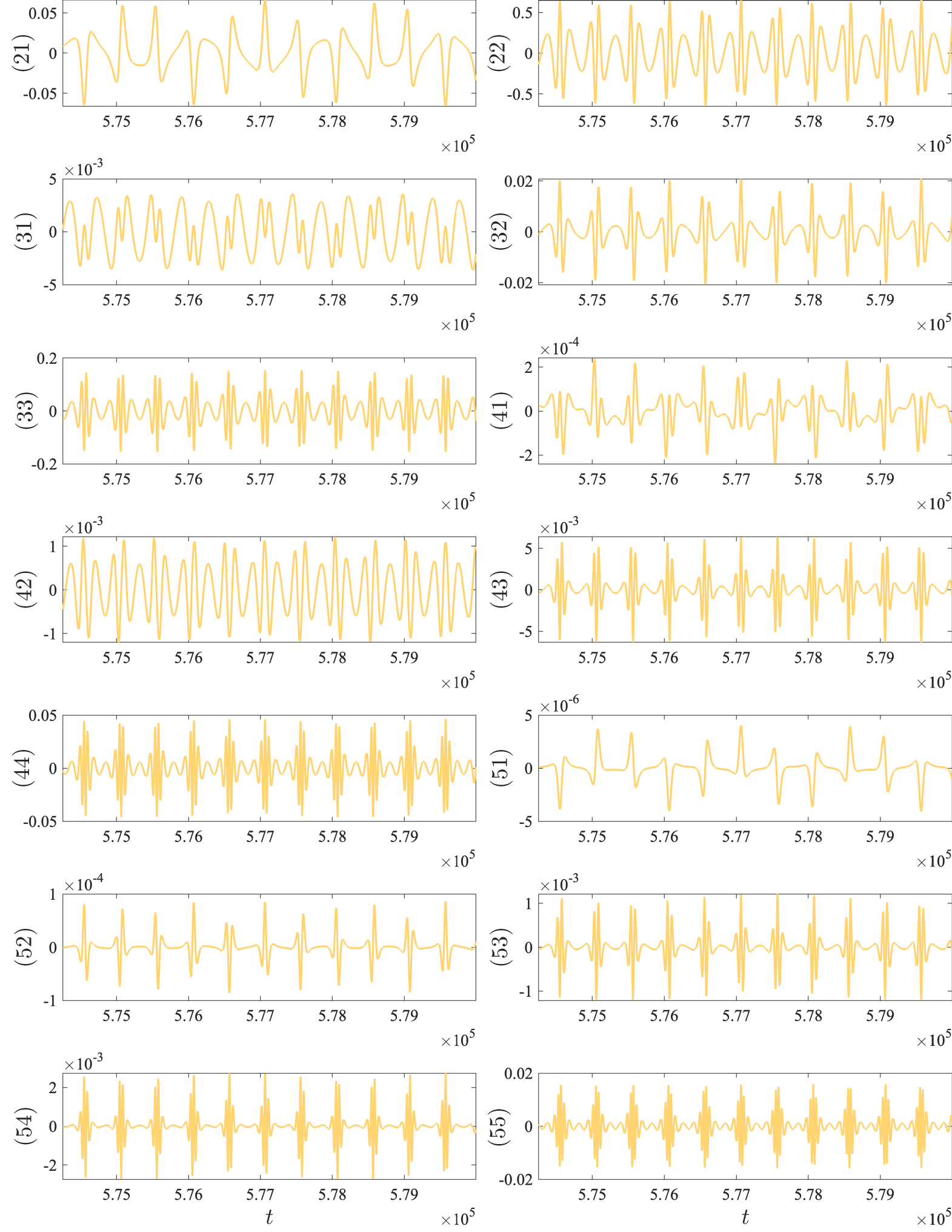}
	\caption{\label{fig:multipoles} Multipoles generated by a binary with $q=10^3$, $e_0=0.5$, 
	$\chi_1 = 0.3$, and $\chi_2=0.1$. The labels of the $y$ axes indicate the values $(\ell \, m)$.
	}
\end{figure}

%=============================
% Figure eccentric spinning, hpc 
%=============================
\begin{figure}[t]
	\includegraphics[width=0.48\textwidth]{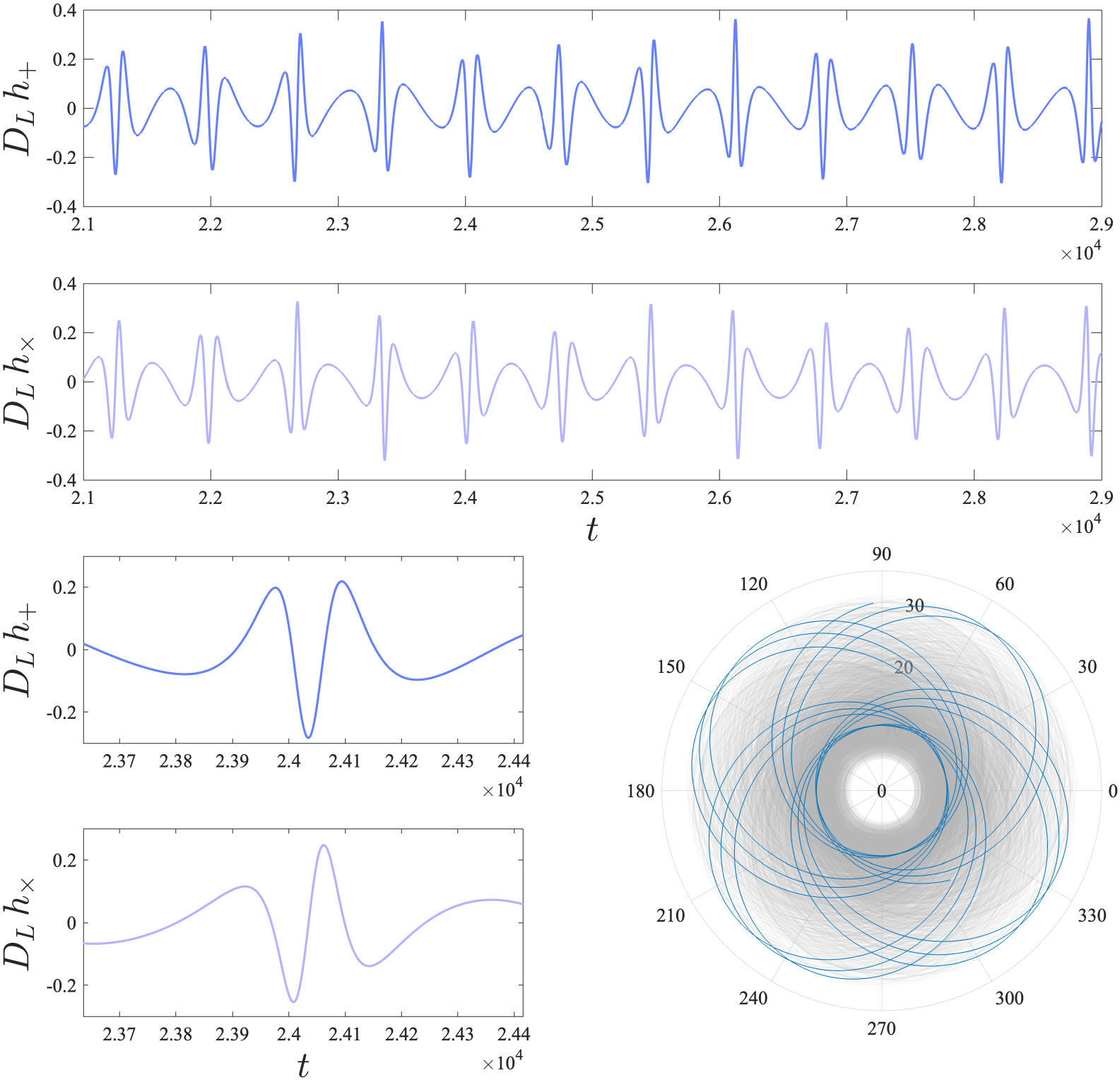}
	\caption{\label{fig:hpc} Strain and trajectory generated by a binary with $q=10^3$, $e_0=0.5$, 
	$\chi_1 = 0.3$, and $\chi_2=0.1$. As for the strain, we sum up to $\ell = 8$ do not 
	include $m=0$ modes, and we consider it as seen by an observer whose line of sight is 
	inclined by $45^\circ$ with the orbital plane. The lower left panels show a zoom on a radial period, and
	the portion of the strain shown in the upper panel corresponds to the trajectory 
	highlighted in blue in the bottom right panel.
	}
\end{figure}

To gain a more astrophysical perspective, let us highlight that the total 
length of the EOB $q = 50 000$ waveform corresponds to $\sim 13$ years for a smaller 
mass\footnote{These values are those considered for the mass of the 
smaller compact object in Ref.~\cite{Babak:2017tow}.} $m_2 = 10 M_\odot$ and to $\sim 40$ years if $m_2 = 30 M_\odot$.
In the former case, the EOB waveform frequency spans the range $[0.0013,0.0096]$ (Hz), while in the latter case
the range is $[0.0004,0.0032]$ (Hz). As LISA will be more sensitive around $10^{-3}$ to $10^{-2}$ Hz, 
we focus on the case for $m_2 = 10 M_\odot$. Instead of taking the full EOB evolution, we choose the 
frequency range $\omega = [0.045, 0.12]$, during which we have $\sim 1.5 \times 10^5$ cycles. 
For the chosen value of $m_2$, the $\omega$ range corresponds to $[0.003,0.007]$ (Hz), and
the associated duration of the EOB waveform is $1.2382$ years.
The accumulated EOB/GSF dephasing on this interval yields $-0.7$ radians (for $\ell_{\rm max} =10$).
This is consistent with the standard requirements of accuracy for EMRIs waveforms~\cite{Lindblom:2008cm, Barausse:2020rsu}, 
since the GSF waveform
(considered as the exact benchmark in this case) is reproduced by the EOB one with a dephasing that is 
below one radian over $\sim 1.5 \times 10^5$ cycles.
These numbers thus prove that, with the options discussed so far, the \Dali{} model can
generate accurate waveforms for EMRIs. 

In the quasi-circular case\footnote{It should be noted that the quasi-circular limit of \Dali{} is slightly different from the native
circular model, \Giotto{} used in the previous section because of the noncircular terms explicitly incorporated 
in the $\ell=m=2$ mode of radiation reaction~\cite{Chiaramello:2020ehz,Nagar:2021gss}. We have explicitly verified that the 
phase difference accumulated because of this latter is $\sim 0.05$~rad, thus much smaller than the current EOB/2GSF difference.
We thus only focus on the \Dali{} model from here on.} (including spins on {\it both} objects), 
the model relies on the iterated post-adiabatic\footnote{The meaning of ``post-adiabatic'' here is different with respect 
to the one used in the rest of the paper, namely not corresponding to an expansion in $\nu$ 
but in $p_{r_*}$~\cite{Damour:2012ky}.} approximation~\cite{Nagar:2018gnk}, that provides iterative, analytical 
solutions to the EOB dynamics ODEs up to a given radius. This technique allows for an efficient generation of long-inspiral waveforms. 
Figure~\ref{fig:benchmark} shows the generation times for $\sim 700$ binaries with total mass $M=10^{7} \Msun$ 
evolved from $0.0001$ Hz with varying mass ratios and spins. When it is employed (top panel), we use the iterated post-adiabatic 
approximation with 8 iterations and stop it at the maximum between the Kerr innermost stable circular orbit (ISCO) and $r=5.5$, at 
which point the system dynamics is solved numerically until $r = 5$ (see Fig.~\ref{fig:rbounds} and Ref.~\cite{Nagar:2018gnk} 
for additional details). Note that we stop the waveform computation at $r=5$ to avoid any contamination related to the
singularity at $r=3$ in the 1GSF-informed potentials~\cite{Nagar:2022fep} due to the gauge choice. In this case, the waveform 
generation time for a typical  EMRI waveform on a single CPU ranges from few hundredth of a second to a few seconds, 
depending on the configuration\footnote{Note that the resulting waveform is obtained with a non uniform time step and an additional 
interpolation on an evenly-spaced grid is not performed.}. By contrast, the waveform generation time obtained simply
solving the ODEs is varying between $\sim 10$~s and $10$~minutes depending on the configuration. 
Focusing on the top panel of Fig.~\ref{fig:benchmark} another few considerations are in order. First, 
up to $\chi_{\rm eff}\sim 0.5$ the generation time is substantially independent of the mass ratio. 
Consistently, it increases with $q$ as long as $\chi_{\rm eff}$ becomes large and positive, because of the
increasingly larger number of cycles of the radiation-reaction driven inspiral that are needed to reach 
$r_{\rm final}=5$. Note however that it decreases slightly as $\chi_{\rm eff}\sim 1$ because the inspiral 
is stopping well before reaching the ISCO.
With the same rational one interprets the bottom panel of Fig.~\ref{fig:benchmark}: the computational efficiency 
decreases with the mass ratio because of the increasingly higher number of cycles for which the system of ODEs
is solved.

As already shown in Ref.~\cite{Nagar:2022fep}, \Dali{} can easily generate waveforms
that incorporate {\it both} eccentricity and spin. As an illustrative example, for a binary with $q=10^3$, 
$e_0=0.5$, $\chi_1 = 0.3$, and $\chi_2=0.1$ Fig.~\ref{fig:multipoles} shows all multipoles 
up to $\ell = 5$, while Fig.~\ref{fig:hpc} reports the total strain as seen by an observer 
whose line of sight is inclined by $45^\circ$ with the orbital plane as well as the orbital motion.
Since for the eccentric case we cannot rely on the iterated post-adiabatic approximation, the computational 
efficiency in this case is expected to be comparable to (or worse than, depending on the configuration) 
the quasi-circular one in the bottom panel of Fig.~\ref{fig:benchmark}.
It must be noted, however, that our focus here is on the physical completeness of the model
and {\it not} on the generation speed of the waveform, that at the moment is the main target of other waveform
generators for EMRIs, like the {\tt FastEMRIWaveform} ({\tt few}) 
package~\cite{Chua:2018woh,Katz:2021yft}.
In particular, we stress that, although we only benchmarked our EMRI model in the quasi-circular,
nonspinning case, many other physical effects are incorporated in the model (notably 1PA effects in the spin-sector and beyond), 
as already described in Ref.~\cite{Nagar:2022fep}. We also note that 1PA information has recently been proved to 
be essential in parameter estimation~\cite{Burke:2023lno}.
Overall, the main scope of this work is to probe the relevance 
of all the elements entering the conservative 
and dissipative sectors, and to prove that the structure of the model is flexible enough
to gradually incorporate everything else that will be needed for physically complete EMRI templates. In this regard, 
we believe  that EOB provides an ideal framework to work in, given the easiness with which we can include more information,
e.g. accretion effects as those described in Ref.~\cite{Speri:2022upm}. 
Let us finally mention that resorting to further analytical approximations, as well as machine learning or reduced-order techniques 
will allow us to obtain more efficient avatars of our model, with significant computational gains and no losses in terms of
physical completeness~\cite{Gamba:2020ljo,Schmidt:2020yuu,Tissino:2022thn}.

\section{Conclusions}
\label{sec:end}
We have shown that the inclusion of  the $\ell=9$ and $\ell=10$ flux multipoles at $3^{+19}$PN accuracy 
in the GSF-informed EOB waveform model for EMRIs of Ref.~\cite{Albertini:2022dmc} is sufficient to 
increase the consistency between EOB and 2GSF waveforms. More precisely, considering a binary
with mass ratio $q \equiv m_1/m_2 = 5 \times 10^4$ and the  dimensionless frequency range 
$(\omega_1,\omega_2)=(0.0224,0.12)$, the additional flux modes reduces the phase
difference $\Delta^{\rm EOBGSF}_{\ell_{\rm max}=8}=3.1$~rad found in  Ref.~\cite{Albertini:2022dmc} 
to $\Delta^{\rm EOBGSF}_{\ell_{\rm max}=10}=-1.1$~rad.
Moreover, when specifying these numbers to the most sensitive part of the LISA bandwith, 
i.e. between $10^{-3}$ to $10^{-2}$ Hz with $m_2=10M_\odot$, we find that the
EOB/GSF dephasing is  $-0.7$~rad for $\sim 1.2$ years of evolution,
which is %\sout{looks}\red{is [SB: we should check and point to a reference]}  
consistent with the standard accuracy requirements for EMRI waveforms~\cite{Lindblom:2008cm}.
It should be noted, however, that \Dali{} naturally includes (or may include) 
effects that are higher order than what is currently included in the 2GSF model. 
In particular, this is the case of the $\nu^2$ and $\nu^3$ (i.e., 2PA and 3PA) 
terms at 4PN, 5PN or 6PN order in the EOB potentials that are known  
analytically~\cite{Damour:2016abl,Bini:2019nra,Bini:2020wpo,Bini:2020nsb,Bini:2020hmy} 
(see e.g. Eqs.~(2), (3), and (5) of Ref.~\cite{Nagar:2021xnh} for the 5PN-accurate potentials). 
These terms {\it are not} included in the 1GSF-informed potentials~\cite{Nagar:2022fep} 
that we are using here, but they may have an impact. We note that analogous  
terms are already included in our $3^{+19}$PN resummed  fluxes. Although 
the influence of these terms is a priori expected to be more important for 
intermediate mass ratios, their importance for EMRIs should be carefully estimated.
Let us finally mention that the spin-sector of the model is currently the one
adopted for comparable mass binaries and might be easily improved with more
spin-dependent information, either in the Hamiltonian (see e.g. the conclusions of Ref.~\cite{Rettegno:2019tzh} 
and Fig.~13 therein for the incorporation of the full analytical information regarding the spinning secondary object) 
or in the analytical flux, that is not sufficiently accurate when the spin of the primary black hole is 
large (see Appendix A and Fig.~15 in Ref.~\cite{Albanesi:2022ywx} for a brief discussion on this matter).
As for the horizon flux for spinning binaries, the current version only includes the leading-order contribution (for both objects), 
but it could be improved for instance by implementing resummed expressions as those of Ref.~\cite{Taracchini:2013wfa}.
All of these updates are doable and will be considered in forthcoming work.
Overall, we stress that the EOB framework is extremely versatile and easily allows for various physical elements to be incorporated,
and that finally speeding up the model will be just a practical matter.

We finally point out that the insight we gained both in Ref.~\cite{Albertini:2022dmc} and here came from a comparison with another formalism.
We are aware that no model can be self-referential, and we are optimistic that various EMRI models will be available in the near future.
We hope for more comparisons to flourish in the next years, and that through these interactions
we will become ready to decode the harmonies the upcoming detectors will allow us to hear.

\acknowledgements
We thank A.~Pound, N.~Warburton, B.~Wardell, L.~Durkan and J.~Miller
for collaboration at an early stage of this work and for sharing with us
the 2GSF waveform data. 
We are grateful to P.~Rettegno for technical assistance during the implementation process
and to S.~Babak for useful discussions.
A.A. has been supported by the fellowship Lumina Quaeruntur No.
LQ100032102 of the Czech Academy of Sciences. 
A.~A. thanks IHES for hospitality, where part of this work was performed.
The present research was partly supported by the
{\it “2021 Balzan Prize for Gravitation: Physical and Astrophysical Aspects”}, 
awarded to Thibault Damour.
R.G.~is supported by the Deutsche Forschungsgemeinschaft (DFG) under Grant No.
406116891 within the Research Training Group RTG 2522/1.
 S. B. acknowledges support by the EU H2020 under European Research Council (ERC) Starting
Grant, no. BinGraSp-714626 and by the EU Horizon under ERC
Consolidator Grant, no. InspiReM-101043372. 
We are also grateful to P. Micca for inspiring suggestions.
Some calculations were performed on the Tullio server at INFN, Torino.

\noindent \TEOBResumS{} is developed open source and publicly available at

{\footnotesize \url{https://bitbucket.org/eob_ihes/teobresums/src/master/}} .

The code version used in this work corresponds to the tag 2310.13578 on the branch dev/DALI-rholm22PN.
The code is interfaced to state-of-art
gravitational-wave data-analysis pipelines: 
\href{https://github.com/matteobreschi/bajes}{bajes}~\cite{Breschi:2021wzr}, \href{https://git.ligo.org/lscsoft/bilby}{bilby}~\cite{Ashton:2018jfp} and \href{https://pycbc.org/}{pycbc}~\cite{Biwer:2018osg}.

\bibliography{refs20240219.bib, local.bib}

\end{document}